# ACCELERATION OF ENERGY DISSIPATION BY BIOLOGICAL SYSTEMS

Adam Moroz, Seng Chong

De Montfort University, Leicester, UK

## ABSTRACT

A general scheme of evolution from autocatalytic processes to socio-technological system is discussed from the perspective of maximum energy dissipation principle. The scheme treats the emergence of biological systems as an initial stage in the acceleration of global free energy dissipation. The sequential emergence of qualitatively new levels of organisation (biological cell, multicellular organism, social system) has been proposed, the levels are suggested as the results of cooperation (symbiosis) of the dissipative systems at the previous levels. The cooperation/symbiosis provides sufficient complexity for development of the next, essentially a new level of energy dissipation, leading to a new form of free energy utilization and a new form of information mapping. From a thermodynamic perspective, every qualitatively new level of biological organisation provides an additional step to increase the rate of energy dissipation from qualitatively new sources, essentially widening the number of these sources involved in the utilization.

## INTRODUCTION

The maximum energy dissipation (MED) principle, together with related maximum entropy production principle [1-4], has been discussed in various fields [3-10]. The maximum energy dissipation principle has been shown to be a good basis for consideration of kinetic non-linearities (cooperativity, autocatalytic~~al~~ growth) in chemical and biochemical reactions and variational decription of dissipative processes [11-15]. As it has been considered in these works, the nonlinearities in processes of energy dissipation are naturally incorporated into the MED principle. This principle can be treated as the general case of the least action (LA) principle, has also the evolutionary implications [15]. On this ground it is reasonable to suggest that the MED and the LA principles are different forms of a principle of least instability, where the free energy can be treated as a quantitative measure of instability.

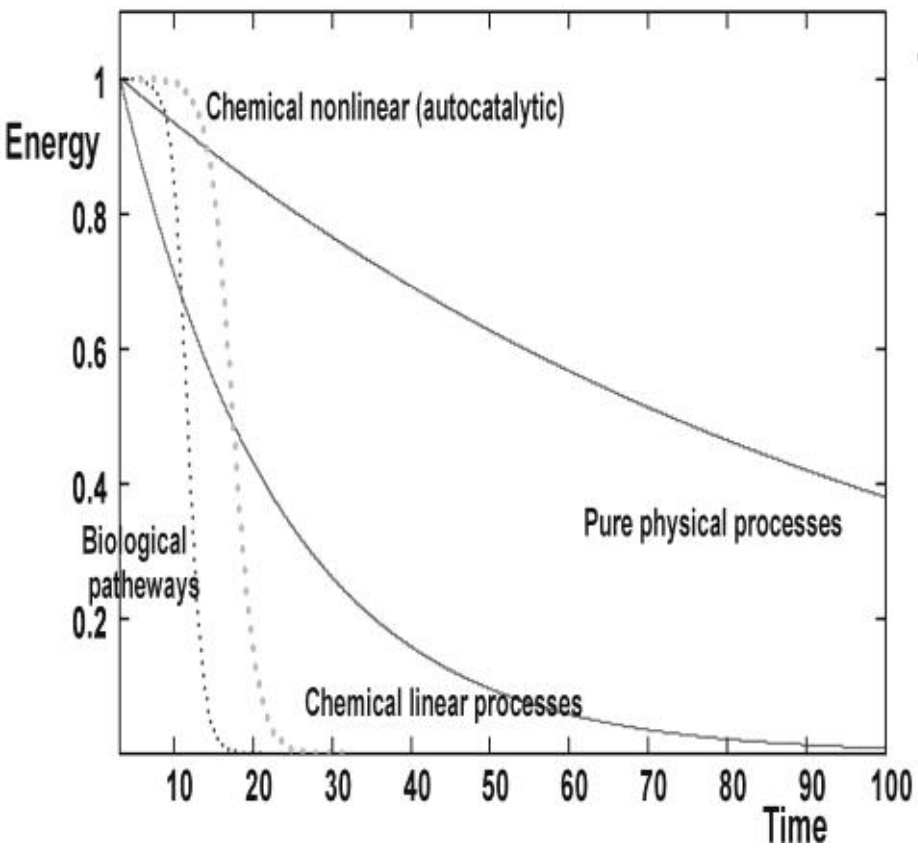

Figure 1. Schematic presentation of the linear and nonlinear dissipative pathways in free energy dissipation. Adopted from [11].

One can see that in such a way, as illustrated in Figure 1, chemical nonlinear and biological pathways utilise free energy more effectively, so the emergence of the nonlinear pathways satisfy the maximum energy dissipation/least action principle. Evolutionary, in a very complex system, having a planetary scale, when the richness of molecular primordial organic soup (diversity of molecular substances) allows, such nonlinear processes can take place. Therefore the overall “dissipative” action (dissipated free energy multiplied by time) is smaller comparably to the linear physical dissipative processes or linear chemical dissipative processes (Figure 1). Then in such interpretation, biological processes are the most effective in sense of the least action (dotted line, Figure 1).

In this work we will discuss some evolutionary and organisational implication of the maximum energy dissipation principle.

## MOLECULAR, PREBIOTIC AND PURE BIOTICAL ORGANISATIONAL-EVOLUTIONAL LEVELS OF BIOLOGICAL DISSIPATIVE PATHWAYS

Considering the molecular level of organisation of biotical processes, it is also reasonable to link them evolutionary to the stage known as molecular evolution. Interpreting molecular organisation in such a way, we have to note that there are two approaches to molecular evolution, in some sense alternative. In a number of works, Russell and coauthors [16] have described an approach when a network of chemical reactions, located and created by a complex environment on the surface of prebiotic Earth, was able to develop a level of complexity, sufficient to generate prebiotic molecular life. Such a molecular network was located nearby the surface and was not separated from the environment [16]. Alternative concept - the Eigen theory of molecular evolution [17-18] is based on known autocatalytic properties of organic polymers (proto-RNA and proto-enzymes). These, in the catalytic sense, are two opposite approaches, which might be combined in the way when at the first, initial stage, the catalytic role of surface prevails to develop a variety of organic substances and later these substances can independently support a hypercyclic network and evolve into proto-cell in a sort of symbiosis with the coacervates. One should note that a biochemical network [19] is a coupled network [20], as of any self-reproductive biological cycle. Particularly in an autocatalytic molecular network, free energy utilization is necessary for synthesis and can be proportional to the rate constants of replication of molecular subspecies involved in such an autocatalytic dissipative processes. The growth of molecular autocatalytic networks is accompanied by utilization of energy rich molecules [17-19] and, therefore, is dissipative. Later in evolution, cellular living organisms, once emerged, had consumed/utilised all resources of free energy reach primordial organic soup, and had developed a spectrum of heterotrophs (mono- and multi-cellular ones) which successfully terminate this soup and everything which was organic which was unprotected and less competitive [21-26].

Evolution of unicellular biological systems went through a number of stages. According to the Margulis endo-symbiosis theory (see, for example, [27]), a proto-eukariotic cell at certain stage has integrated a chloroplastic cell and proto-mitochondrion. Mitochondrion is known as a semi-autonomic subcellular organelle with its own two-strand cyclic DNA, indicating the bacterial origin and similar to bacterial mechanism of transcription/translation.

Based on the assumption that cooperation of the same level biosystems, as sort "dissipative autonomic agents" of similar level of organisation (cellular systems) can provide additional adoptivity for the species and opportunity to develop a new dissipative degrees of freedom, necessary for surviving, an evolutionary transition from the single cell organisation to the social pinnacle can be built as in [15].

Then the overall scheme can be expected as a series of levels with a superordination, superinclusion and coevolution. The qualitative evolutionary transitions, as are seen from this scheme, can be characterized by the following qualititative transitions [15]:

+compartmentalisation of macromolecules with hypercyclically-like auto-catalytic properties which evolved into a proto-cell developing a number of catalytic and informational molecular processes;

+forming a symbiosis of some proto-eukaryotic cells and their subsequent evolution into proto-multicellular organism;

+formation of social super-organisms by some biological species;

+emergence in the framework of social systems, a symbiotic relationship within some nonbiotic things that essentially extends functional and adaptive abilities.

The organisational structure, related to the evolution of the free energy consumption/dissipation by biological processes can be schematically represented by modifying so-called trophic pyramid illustrating the organisational hierarchy of biological systems (Figure 2).

Molecular autocatalytic networks can be considered as a first stage, an initial level of organisation of dissipative processes (Figure 2). However, just those molecular subspecies survived (and gave the life for a protocell) which were capable to develop a protection from environment coacervate-like encapsulation – membrane and cellular wall. So the second stage, second level can be linked to the organisation forming a prokaryotic-like cell (Figure 2).

The third level in the scheme (Figure 2), is introduced as a whole spectrum of eukaryotic cells, more precisely, spectrum of unicellular species. Evolutionary just a certain part of these species (designated as "cellular species capable to form multicellular organisation") were able to form a multicellular form of organisation. Not all unicellular species had the capability to develop the next level of biological organisation by cooperation. One can note that the multicellular organism has been developed as a result of a long evolutionary process.

The fourth level can be represented as a level of multicellular organisation, where the cells are forming an organism characterised by the integration and specialization of the activity of all cells. This resulting integral activity cannot be considered in terms of a single cell. Even for prokaryotic cells, the cooperation between cells is widely observed [28]. Modern multicellular organisms represent the biosystems evolved throughout millions years. These organisms formed a different from the cellular metabolism, regulation, cognition and can be considered as next level in Figure 2. This organisational level can be represented by a number of multicellular species, which have developed an essentially new degree of freedom of competition as, for example, the locomotions (running, swimming, jumping, flying). The ability to move fast provided multicellular organisms with an important method to find food, to escape danger, to develop also the new integrative for this level informational cognition – eye-seeing, hearing, brain.

The fifth level can be considered as a level of those multicellular systems which were capable of forming so-called superorganisms – sometimes also referred to as communities or families of individuals. Such a known species as ants, bees, termites can be good examples of social species [29-30], organised in colonies, superorganisms. This kind of biological organisation can be characterised by the mode when the needs of superorganism/colony have a priority comparably to the individual needs. Such a sort of organisation provided an adaptive and competitive advantages for these species. Their social organisation is characterised by partial usage in their activity of things of non-biotical origin, which extends functional abilities of individuals and adaptation of the superorganism.

However, from the spectrum of social species just *Homo sapience* (HS) was capable to "spin-off" the non-biotic origin means of production to the stage that they achieved self-reproductive-like properties. In biological terms, HS was able to form symbiotic-like relations by means of production, see also next sections. This provided to the HS social system tremendous opportunity to explore a wide number of free energy sources not accessible or fragmentary accessed by other species. The sixth level can be considered as the level of social organisation widely exploring technology.

Following proposed scheme, the bioevolution is a result of the demand of maximal free energy dissipation: the emergence of every qualitatively new level of bio-organisation is due to capability to acceleration of dissipation of essentially new qualitatively sources of free energy.

The scheme in Figure 2 incorporates the minimal evolutionary mechanism: throughout the cooperation of (in cybernetic terms) dissipative autonomic agents to the formation of essentially new form of organisation, which allows utilization of new free energy from formerly inaccessible free energy sources; which can be seen at every evolutionary change on every level of hierarchy - cellular, organismic and social.

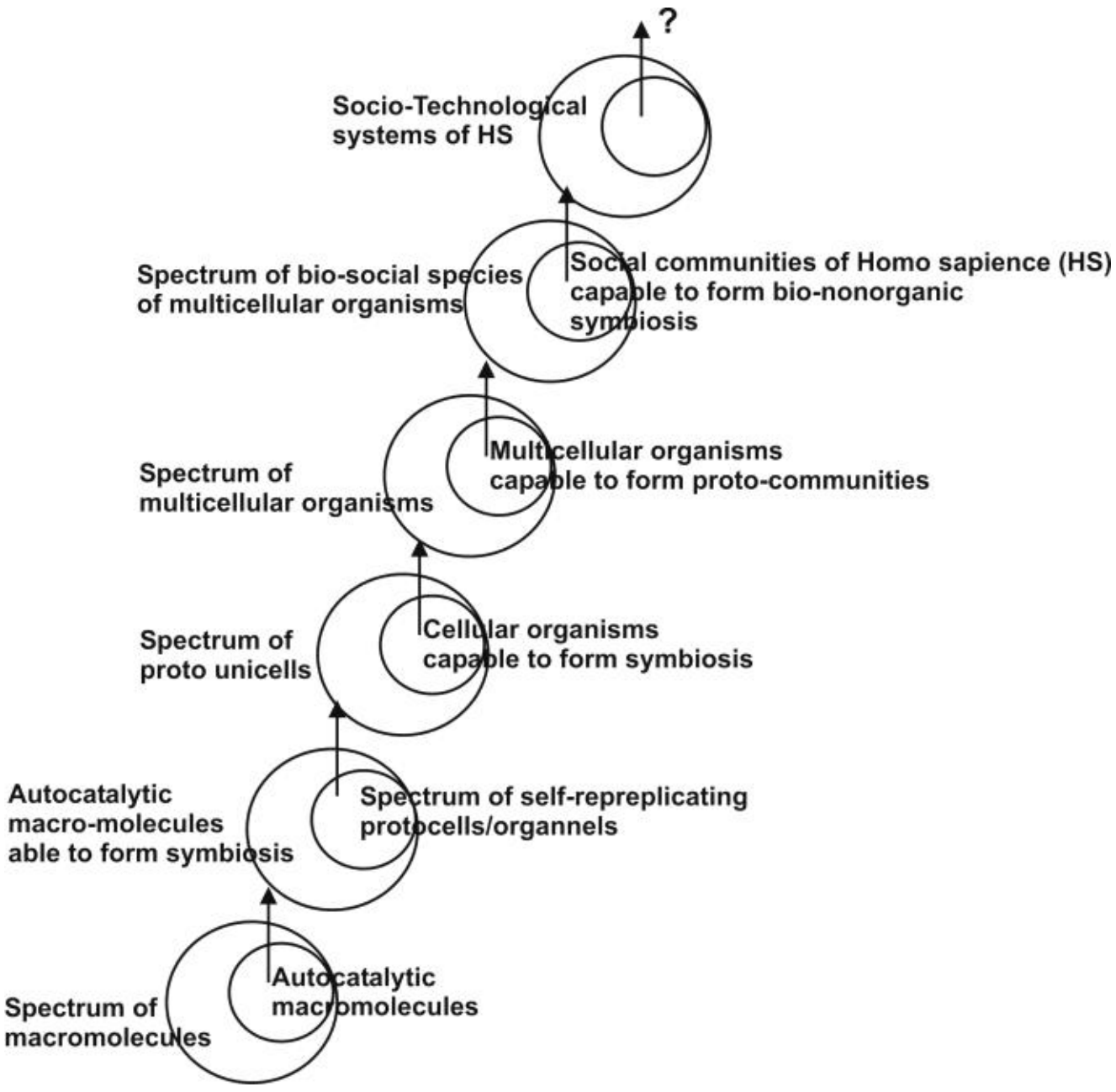

Figure 2 Evolutional ladder of dissipative systems. Schematic levelled representation of the organisation and evolution of dissipative cycles: from molecular Hypercycle to technological Supercycle (adapted from [15]). The role of symbiosis/cooperation during qualitative transitions in the trophic pyramid of dissipative systems, including bio- and biosocial-systems. The ability to utilise essentially new forms of free energy is related to the transition to the next step of the cooperative interaction of the biological systems at every level. The main outcome of symbiotic interaction is the formation of a qualitatively new type of integration and differentiation of the functions in the formed cooperative system, and also the formation of a qualitatively new form of the information mapping and a qualitatively new way of dissipative transformation of free energy.

An important note is that the cooperation between the autonomic dissipative agent of any bio-level provides not just better ability in competition, but also a potential to develop a qualitatively new level of organisation, qualitatively new level free energy consumption (dissipation) and qualitatively new level of information processes (cognitive), supporting dissipation.

From a thermodynamic perspective, bioevolution is a dissipative coupled process accelerating overall utilization/ dissipation of free energy. From this perspective the reasons of emergence of biotical dissipative pathways - biological systems are purely physical – so physics and MED principle demand emergence and evolution of biosystems. Biosystems cooperate for the increase of adaptivity and from the thermodynamic perspective it helps to develop consumption of different sources of free energy, not accessible from the previous level of organisation. The utilization of new energy sources initiates a divergent phase in the development of species.

Thus, biological phenomena are the extreme phenomena in the sense of energy dissipation, they utilize free energy from sources where somehow usual physical mechanisms do not work or work insufficiently fast. Since the maximum energy dissipation principle demands fastest possible dissipation, biological and socio-technological phenomena satisfy this demand, and their emergence and existence are consistent with MED/LA principles and whole physics.

Previously [15] it has been concluded that each qualitatively new level of biological organisation becomes possible due to the cooperation/symbiosis of structures of the previous level, further specialization and integration of these structures within the framework of the association emerged, which developed essentially new form of substance-and free energy utilization and essentially new form of information cognition.

However, it can be seen that every level of biological organisation has its own limitations. These limitations are set up by nature and the scale of free energy utilization and material consumption as well as in the nature informational support of metabolism, which characterizes the particular level [15].

The level of pre-biotical processes (which has vanished in early stages of evolution and in modern biotical world can be thought as molecular network level) was limited to utilization of free energy-rich chemical substances available in a primordial organic soup. The metabolic processes obviously were limited by utilization of the limited range of energy rich molecular substances.

Emerged from autocatalytic networks/cycles the unicellular organisms in a long evolution accompanied by few symbiotic events explored a wider number of free energy sources. However, the unicellular level of biological organisation is limited in scale and nature of energy sources utilizing [26].

One can also see that the biomass, in fact free energy in a biotic material form, produced by multicellular biological systems, has the scale of billions tones, but still is limited and has a order much less than 1% of solar radiation in energy equivalent. Therefore, one can conclude that all biological levels and as we will illustrate below, the socio-technological processes, are limited in the scale of free energy consumption [15].

## ACCELERATING DISSIPATION BY SOCIO-TECHNOLOGICAL CYCLE

Above, the hypothetical scheme, Figure 2, of the emergence of biological systems, their organisation and evolution as the dissipative systems, based on MED principle has been discussed, following [15]. The result of their evolution is the sort of organisational ladder from biopolymer macromolecules through their networks and cellular and multicellular organisms to the top level - social superorganisms. The key points of these formations are the cooperation between the system of same level to constitute initially nonintegrated associations and later to develop and evolve to a new level of systems, with essentially/qualitatively new levels of free energy processing and informational mapping.

Such a sort of cooperation (e.g. social form of cooperation, social symbiosis) widens the adaptivity of such a community (local population), increases the territorial competitiveness for food, and such a community has improved chances to survive. This trophic aspect can be interpreted in the thermodynamic sense, because biomass is a type of free energy and its consumption therefore is a dissipative process. Thus, from a thermodynamic perspective biological species are just specific dissipative processes (generalized biological flows) that over-shunt, overtake, develop and compete with each other for free energy resources. In that sense the considerations from the maximum energy dissipation can be applied [15].

At the top level of these organisational developments, the cooperation produces the social level of organisation, known for a part of HS for several other species. Classical examples are superorganisms - formed by ants, bees, termites. However a long organisational distance divides them from humans. Would be it possible to explain this distance from the perspective of energy dissipation and the main mechanism of developing such a new level of organisation – cooperation or, biologically speaking, – symbiosis?

Let us note the important observable difference: the cooperation between the biosystems at the same level of biological organisation (macromolecules/networks, cells, individuals) always was carried out within the systems of biological nature exclusively. The observation of the humans' cooperation can be considered as the cooperation between biological individuals (humans) and the tools/means of production, having *non-biological* nature [15]. This in fact can be pointed out as the key difference between the social system, formed by HS and all others pure biotical social systems: one can say that humans invented technology.

Starting from a simple use of primitive tools, humans developed very complicated technological processes, which, one can see, indicate the self replicating-like properties. This helped humans (in the framework of developed new form of superorganism –society with technology) in thermodynamic terms to extend the overall process of energy dissipation to a very wide number of non-biological sources of free energy.

From the perspective of dissipation, only socio-technological system has jumped from the utilization of only biotical free energy sources, characteristic for bio-systems, as well as for known social species, known in the biological world. Only HS socio-technological system expanded usage free energy to qualitatively new sources and developed an essentially new level of organisation of free energy processing pathways. This utilization of free energy obtained a global character. Also, the essentially new form of informational support/cognition has been evolved in the process of evolution of the HS free energy processing and the development of the means of production accelerating pathway. These new pathways nowadays are dominating in energy consumption (dissipation) by the HS socio-technological system, Figure 6.

The self-reproductive-like, symbiotic interaction starts as pre-historical usage by HS individuals of the non-biotical origin tools which were able to develop more and more complex and useful ones. Throughout thousand years these tools have been successfully modified. One can see that from a self-organisational perspective, a minimal scheme of this evolutionary mechanism can be similar as for cells, which cooperate into a cell colony and evolve into a multicellular organism, [15].

As a result of such a "bio-nonbio" symbiosis, both components- biotical (biomass, the overall population of HS) and non-biotical (the means of production) grow tremendously. One can note that for the growth of biotical component a few critical stages observed. They can be linked to few key developments in exploration of new resources or the development of non-biotical "means of production" component (see, for example [15]). The first can be related to agricultural phase and another one in linked to technological phase.

The nonlinear growth of the nonbiotic component (means of production), having self-reproductive-like properties in biological terms, is rather characteristic of the industrial age; its self-reproductive-like growth is discussed in [15]. The main indices/parameters of technological, non-biological components also indicate the exponential growth (see the statistical data from US Census [37] or data from the Angus Maddison's site and works [38-40]), which is similar to self-reproductive growth kinetics of biological systems [15]. One can conclude that this sort of growth supports the suggestion of indirect self-reproductivity. This sort of growth especially is characteristic for different types of energy produced in different economies. The various economic data indicate exponential growth for many indices. This proves the acceleration of the energy usage and dissipation by the socio-technological system, which can be illustrated as additional impact into global energy dissipation rate, Figure 4.

So, one can also note, as in [15], that every level in the trophic pyramid organisation, Figure 2, can be characterized by the limitations in the value of involved and consumed free energy. Single cell and multicellular organism are limited by quantitative and qualitative variety of free energy utilised. The socio-technological system is as well limited in the utilization of free energy.

Thus, one can see also the energy utilization limitations both - in purely biotical and bio-socio-technological parts of the global trophic pyramid (Figure 3). As a summary, in biological trophic (dissipative) pyramid one can observe several qualitatively essentially different levels of organisation of structural-energy dissipative transformation of the free energy flow from different sources (see also [15]):

+level of pure physical dissipative processes

+level of prebiotic, molecular evolution processes,

+level of purely biological processes with unicellular organisation,

+level of organismic organisation or multicellular organisms,

+level of social species organisation evolving to socio-technological system of HS.

These levels are also limited in their information mapping/cognition of the environment and information, necessary for self-replication in a wide sense.

## ON POSSIBLE POST-SOCIAL STAGES OF DISSIPATIVE SYSTEMS EVOLUTION

The scheme in Figure 4 and related considerations imply that the development of hierarchy of biological global dissipative processes expands toward the exploration of new resources of free energy, employing the cooperation and further specialisation and integration of processes at every level. As the result, the essentially new levels of processing of structural-and-material forms of free energy and the essentially new levels of information processing are formed and evolved, which make the hierarchy of biological processes completed and then the essentially new technological level emerged. Then, it is reasonable to suggest from observed biological and socio-technological parts of global free energy dissipative pyramid, that a next level of cooperation between technological systems has to be considered. As a possible result of forming such a new level of energy processing/dissipation and evolution, as one can suggest, can be a space/cosmic organisation of human-like socio-technological systems, which is continuing forming their technological activities at solar system toward to an unseeable scale. Then the human socio-technology can be considered as an elementary subsystem in an organized super technological system, similar to a single cell in a multicellular organism or an individual organism in the biotical societies.

## SOME GENERALISATIONS

Presented above scheme of evolution of organisation of biological systems is based on the suggestion that driving force of the chemical, probiotical, biological and socio-technological evolution is the maximum energy dissipation principle, which can be treated as a partial case of the least action principle.

We expect that an evolutionary role for this fundamental principle for all physics: maximum energy dissipation/least action principle can be considered as an evolutionary principle, stating that those dissipative processes win in evolutionary competition, which could provide the highest possible rate of free energy dissipation. Kinetically, the maximum energy dissipation principle is a principle that employs nonlinearities.

Indeed, the maximum energy dissipation principle in combination with the maximum entropy production principle is probably the only physical principle that can explain the emergence of biological systems as the complex acceleration forms of free energy utilization. The relationship of the maximum energy dissipation principle to the least action principle unites, connects the biological processes to physical processes. In fact, these principles make the biological processes so universal, as physical processes. Thus, unified LA/MED principle can explain the emergence of biological pathways of dissipation. Moreover this unified principle can explain also the stages of the biological evolution and its transition to socio-technological evolution.

In the energy transformation performing by biosystems, one can note few key limitations, which can be linked to the thermodynamics properties and MED principle. First one is related to the variety and scale of the structural/material forms of free energy utilization/dissipation. Second limitation is linked to information (negentropy generation): the informational support of maintaining/providing the free energy utilization by biological systems is essentially limited in the mapping capability because of a material nature of informational subsystem.

Thus, the MED principle implies that the emergence of biological systems is a physical process. Indeed, the thermodynamic perspective can play a vital role in consideration any endogenous way of life emergence on our planet. Particularly the MED principle welcomes the emergence of the autocatalysis from the huge spectrum of catalytic processes.

Due to the diversity of chemical processes, the number of possible chemical dissipative pathways over-exceeds the number of several physical processes of energy dissipation by the number of decimal orders. Among them there is a number known of different catalytic processes. From MED perspective is not particularly important how the self-replicative molecular structures initially emerges, where they were synthesized - on the surface of the prebiotic planet (Russell and coauthors [16]), or they emerged as the result of self-polymerization in a complex network, proposed by Eigen and coauthors [17-18]. Important is that they bring the nonlinear, faster way of free energy dissipation, as illustrated in Figure 1. Significant is that they are capable to nonlinear acceleration of free energy dissipation - free energy in a form of various molecular compounds/structures. Certainly, that the physical and chemical processes of energy dissipation coexist. Their competitiveness has an indirect form and is rather related to their existence.

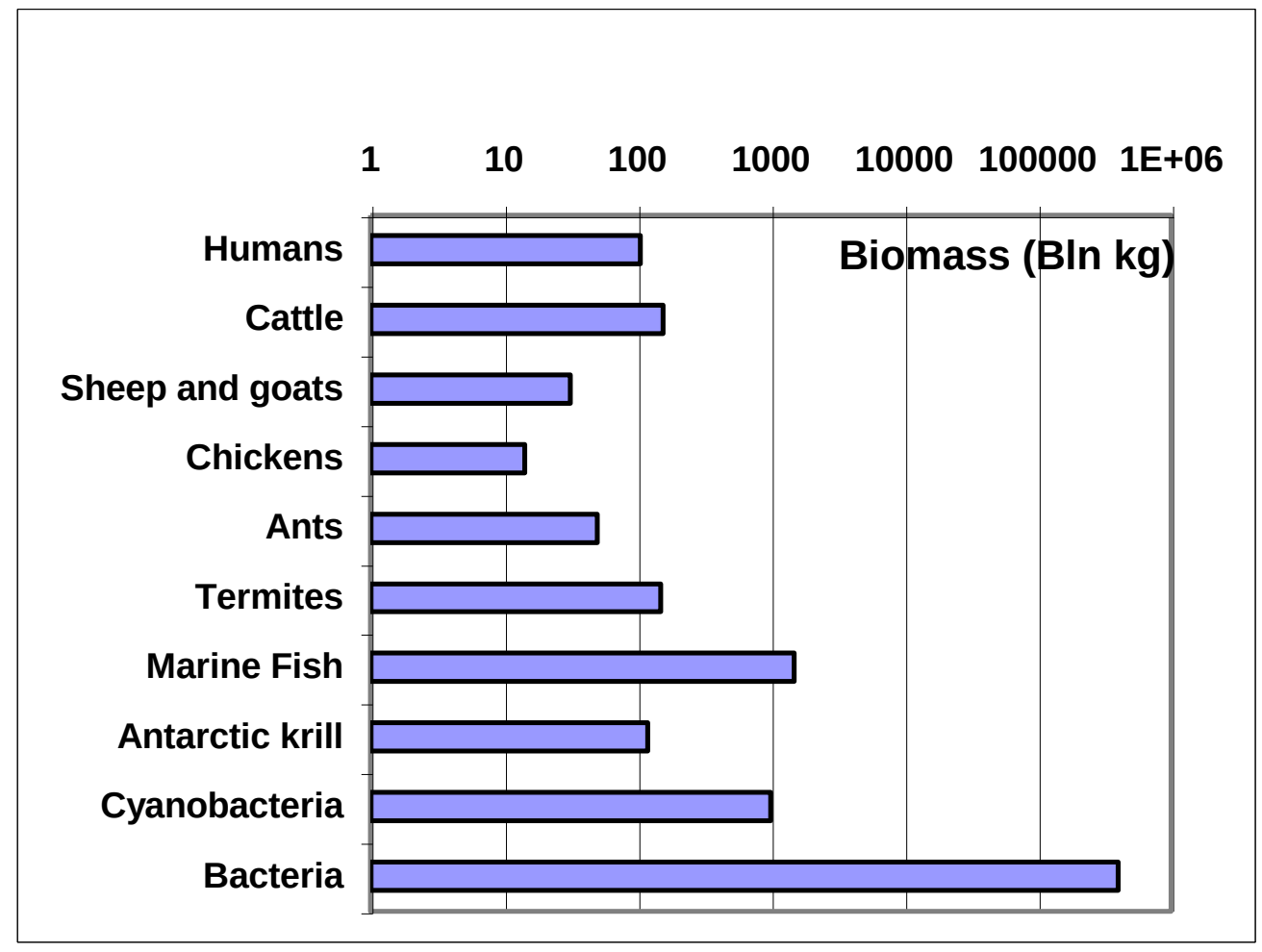

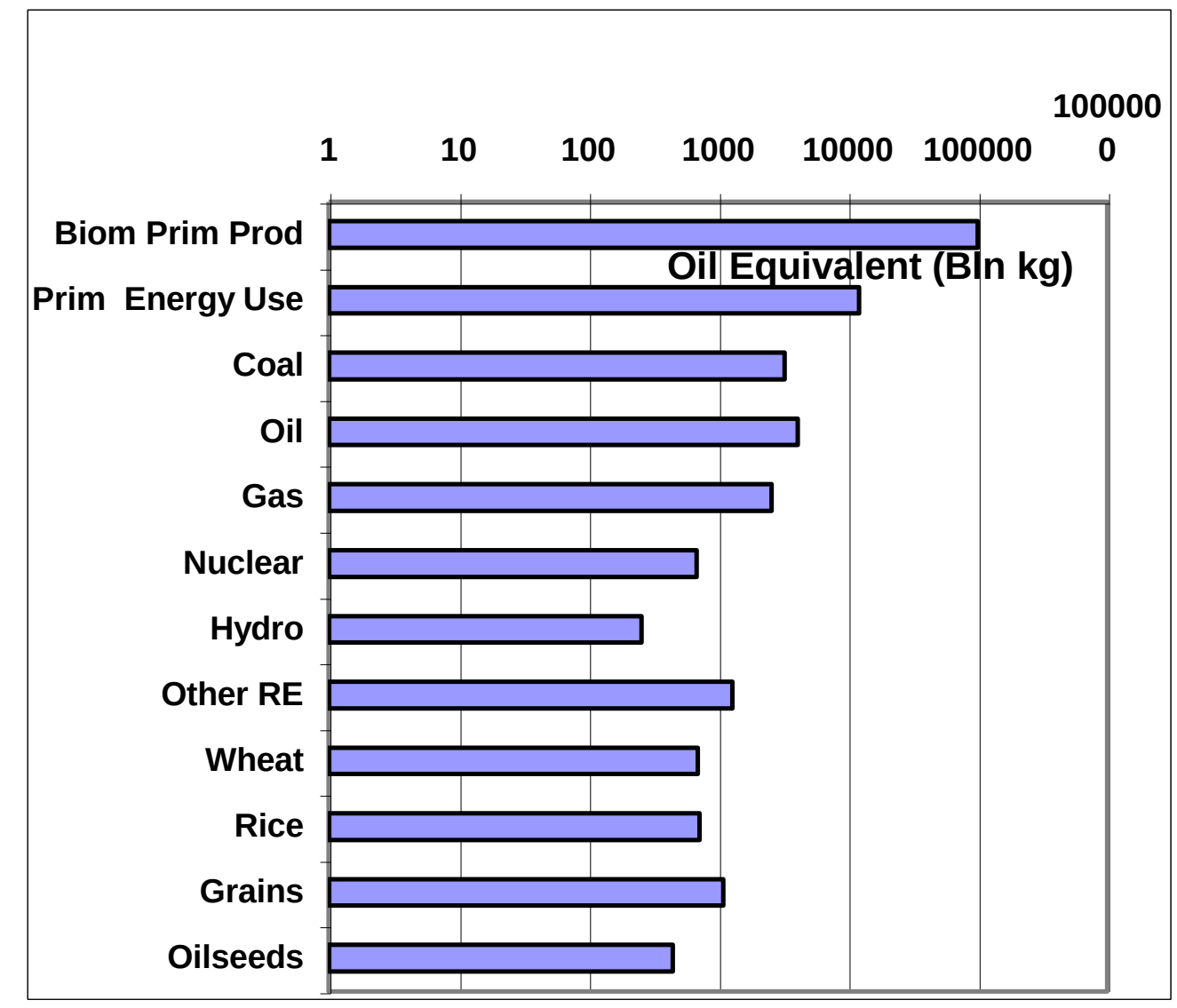

Figure 3. Global scales of energy utilization by some biospecies and socio-technological system, based on [31-40].

On the other hand, the gain, the acceleration of free energy dissipation by nonlinear autocatalytic processes can be considered as the emergence of non-pure-physical degrees of dissipations. The pure physical nonlinear dissipative processes can be observed in stimulated emission of the gain medium in the lasers. The nonlinear relaxation takes place and it has a pure physical nature. In more natural environment, the MED principle welcomes the non-linear processes of chemical nature, which may occur at the surfaces, e.g. catalytic processes. Kinetically, all these processes can be nonlinear, having at initial stage an exponential growth. In a spectrum of macromolecules, which can be initially randomly synthesized on the surface of prebiotic Earth, later the autocatalytic macromolecular structures emerged. Particularly these molecular structures can start the emergence of molecular networks and the hierarchy, which can lead to the emergence of prebiotic Eigen hypercycles-like molecular networks, having self-replication properties and RNA-based informational molecular cognition.

The cooperation in different forms has taken place in earlier stages of the emergence of biological systems. It has taken place in the form of molecular symbiosis. In the framework of this cooperation/symbiosis, further developments of molecular forms of informational accumulation for more effective metabolic networks or primary structure of macromolecules, secondary structure, enzyme activity, informational control of functionality, functional/enzymic support of spatial separation from the environment have taken place and led to the formation of prokaryotic-like proto-cell.

The overall trend and qualitative transitions in the evolutionary process of biological systems can be explained by the complexity increasing based on symbiosis/cooperation as a universal mechanism to develop a new qualitative level of free energy consumption/dissipation. The emergence and formation of these symbiotic associations opened the capability to expand to a new dissipation level. New functions are gained by the capability of such a symbiotic organism to further develop the qualitatively new organisational structures and the exploration of qualitatively new sources of free energy.

The hypothetical general scheme in Figure 2 illustrates an overall trend in the biological hierarchical organisation. In some sense these levels of organisation are also the evolutionary levels, levels of major transition in the form of dissipative relations with the environment as a source of free energy. According to the maximum energy dissipation principle the key characteristic of biological evolution can be considered as in [15]:

* every new level of organisation/evolution (cellular, multicellular, biosocial) emerges as a result of cooperative, symbiotic relationships, further specialisation and integration in associations/communities formed by previous level of organisation biological systems;

* every major level of the organisation can be characterized by essentially new level of dissipation of the forms of free energy resources which are not possible to utilise at the previous organisational levels of free energy dissipation. This includes also new biological resources which appeared as a result of the process of evolution. Overall acceleration of dissipation can be schematically illustrated in Figure 4. Every transition in organisation indicated in Figure 2, is accompanied by increase of global rate of energy dissipation.

* the initial phase of the emergence of a new possible dissipative level is accomplished by divergent phase of the spectrum development, when emerged new structures develop various forms of organisation and the overall number of species significantly increases;

* the competition between different dissipative pathways increases when the total dissipation rate reaches the scale of free energy inflow into the system. Then, in biological terms, the integrative functionality of biosystems becomes crucial for the competition, specialisation of the constituent subsystem becomes very important and, as the final result, the essentially new level of integrative functionality and development of a new form of informational/cognitive support finally leads a new level of organisation;

* the cooperation between the same level of organisation biosystems in the whole spectrum of species provides a capability to develop new level of symbiotic structures as an essentially new manner to adapt and survive. From a dissipative perspective, that lets to final development of a new way to dissipate free energy in an accelerated way, including exploring the new forms of it. Indeed, the cooperation, plays a crucial role in the emergence of new organisation, new biological systems;

* at every level of biological organisation, there are qualitatively different informational sub-systems providing informational/cognitive support for the optimization of energy use/exploration and overall competitiveness of the biological structures at this level.

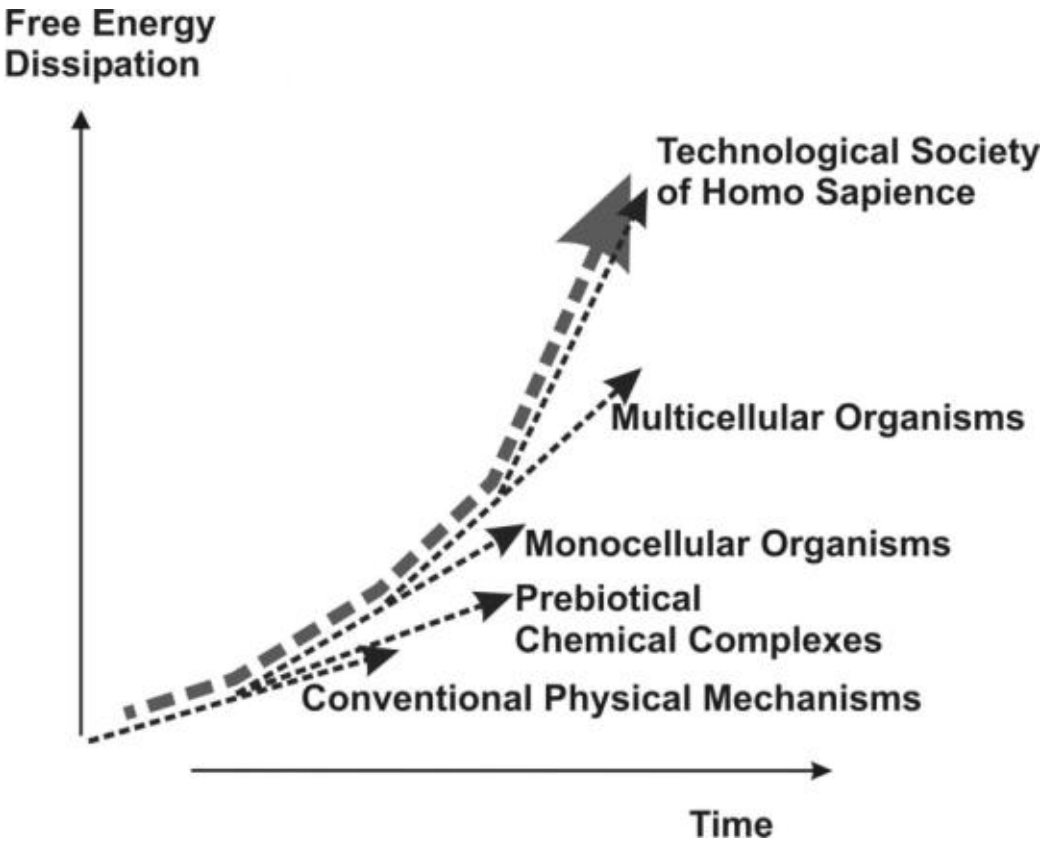

Figure 4. Schematic presentation of acceleration of energy dissipation at every stage of evolution of biosystems.

Finally, one can also suggest that by forming certain basic and minimal evolutionary step-unit, Figure 5, discussed also formerly in [15] can be considered:

*cooperation (symbiosis) of the dissipative processes at level-down (previous level)

*development of a qualitatively and essentially new form of organisational processes in the framework of new type of cooperation/symbiosis,

*extension/exploration and utilization (or faster utilization than at previous organisational levels) of new free energy sources as a result from dissipative perspective.

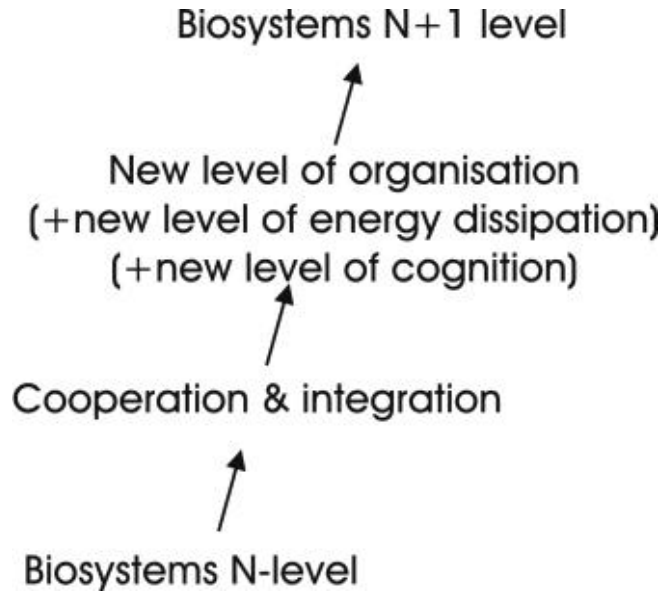

Figure 5. Principal evolutionary step, characteristic for "evolutionary ladder", see Figure 2.

Using this minimal evolutionary mechanism/step, Figure 5, it is possible to explain the transitions/evolution of bio-systems at every qualitatively new level of biological hierarchy – cellular, organismic, social and even socio-technological. In this way, the biological life, its emergence as a phenomenon characterising by its low probability, is a very robust and probabilistic process, and its emergence, robustness and evolution towards complexity is provided by MED principle.

Taking into account discussed above, can be concluded that the emergence and evolution of socio-technological dissipative pathway can be seen as a result of evolutionary aspect of the MED principle. As it is shown in the general scheme, Figure 2, the socio-technological level of organisation predictably emerges/evolves from biosociality, which is characteristic of a number of biological species. However, just HS species developed a global socio-system, which has complex organisation in any aspect - social, energy and substance processing and informational processing. As we have noticed above, the socio-technological system is the unique product of cooperation of biological component (individuals and the labour) with non-biotical component - the means of production, having non-biological nature. In terms of energy dissipation, socio-technological system of HS considerably expands the pure biological scale of free energy exploration. Its emergence and evolution corresponds to the maximum energy dissipation demand or the least action principle.

This socio-technological level can be considered as a top of the biological trophic pyramid. It is essentially a new bio-socio-technological level of organisation which can be treated as a symbolically-like related to certain non-biotical functional objects having a non-biological origin and characteristics (which usually called as the means of production). This is a key difference of HS "social organism" comparably to other known biological social species, like ants or termites. The data on biomass of these social species is indicated in Figure 3. One can note that the biomass of these species is very big, which can indicate the powerful role of social way of organisation in the adaptation of biological species to environment. However, only the HS social system was able to develop symbiosis-like relations with a non-biological means (later in evolution- the means of production), having self-reproductive-like properties. Due to these symbiotic interactions of Homo sapiens with nonbiotical structures, the Homo sapiens social system was able to extend energy-structural consumption of qualitatively new energy resources of non-biological origin [15], and has developed qualitatively new mechanisms of energy processing and qualitatively new information processing (cognition). In this way, the socio-economic evolution appears as a continuation of bio-social evolution.

In Figure 6, the approximate scales of informational mapping, which can be considered as more sophisticated at a higher level of biosystems (like multicellular or social), as are indicated. One can see the expansion of this mapping with the increase of organisational complexity in direction to socio-technological system. However, one can expect also the limitations at every level of biological organisation [15].

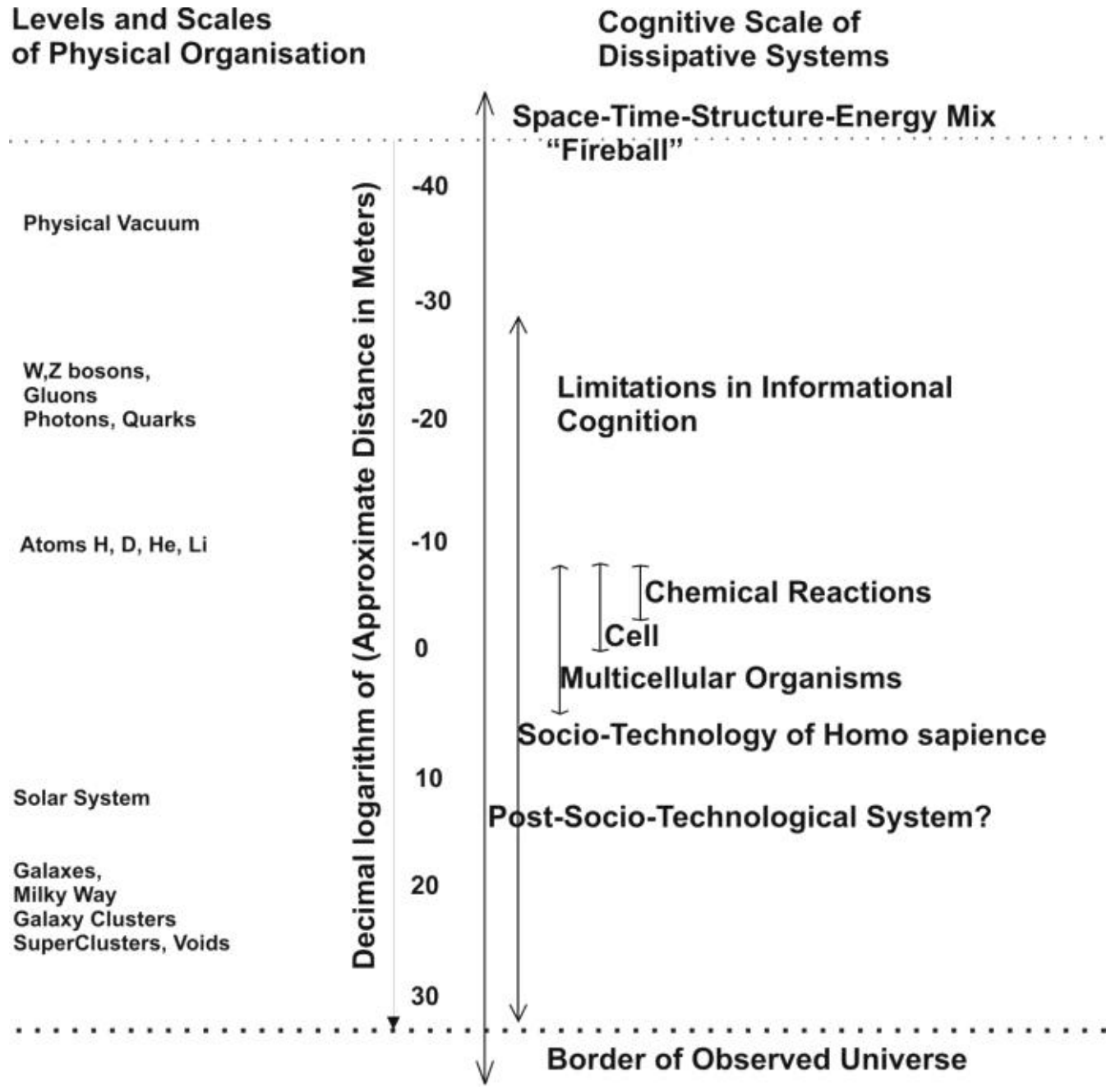

Figure 6. Schematic presentation of scales of informational cognition of the systems of different levels of organisation (adapted from [15]).

However, it is expected, that free energy processing scale and informational support in socio-technological system indicate their fundamental limitations [15]. First limitation is directly linked to the scale of exploration and processing of energy, which is limited by technological capability. Second limitation – is linked to socio-informational mapping/ cognition – that part of technology which deal with the investigation of new energy sources (for example, nuclear fusion) and develop the processes of energy production in conventional forms suitable for standard usage in technology and for effective and qualitative consumer goods production.

Based on evolutionary trend, proposed in Figure 2, a post-socio-technological organisational level of dissipative processes can be expected as a next organisational level after the socio-technological organisation systems. This new level can be characterized by qualitatively new level of energy consumption and linked to it dissipative process carried out, and also a qualitatively new level of information mapping/cognition. Taking into account cognitive limitations at every biological level [15], illustrated in Figure 6, one can state a question: to what extend is it possible to predict the characteristics of this new post-socio-technological level. Is it possible to forecast within the framework of limited socio-technological and informative-scientific mapping of Homo sapiens?

Taking into account that in proposed evolutionary mechanism, every qualitative new level of dissipative systems organisation always emerges as a result of cooperativity of the systems of lower/previous level, one can suggest that a higher to socio-technological level can be the level when number of socio-technological systems cooperate, interact, specialize and form even more complex organisation, [15]. One can expect that the scale of this organisation can be enormously large. Then such extra-terrestrial expansion of HS-like technological systems can be considered as a first stage in the emergence of this post-socio-technological level of organisation. However, an expansion of post-socio-technological system into ~~a~~the small distances of physical world can also take place, see Figure 6. This is expected to be a combined expansion in both super-large and super-small, plank-scale physical worlds.

## SUMMARY

It is suggested that the driving force for the molecular, prebiotic, biological and socio-technological evolution can be explained on the basis of the maximum energy dissipation principle, which can be treated as a partial case of the least action principle.

On this basis, a general scheme of evolution from autocatalytic molecular processes to socio-technologic system can be built from the perspective of maximum energy dissipation. This scheme also reflects the organisational structure of biological world. The scheme treats the emergence of biological systems as an initial stage in the acceleration of global free energy dissipation. The sequential emergence of qualitatively new levels of dissipation (biological cell, multicellular organism, social and socio-technological systems) has been discussed, and the levels are suggested as evolved in the results of cooperation/symbiosis of the dissipative systems at the previous levels. The cooperation/symbiosis provides the basis for the development of every next, essentially new level of energy dissipation, leading to a new form of free energy utilization and a new form of informational mapping/cognition. From a thermodynamic perspective, every qualitatively new level of biological organisation provides additional step to increase the global rate of energy dissipation from qualitatively new sources, essentially widening the number of these sources involved in the utilization. In the sense of maximum energy dissipation/least action principle, the emergence and existence of biological and post-biological (socio-technological) systems are the direct requirement of this united principle, which makes the existence of the biological world in agreement with physics as a whole.

Post-socio-technological level can be considered as the cooperative, symbiotic-like organisation of socio-technological systems as the subsystems, elements. This level can be characterized by significantly new level of energetical and substance/matter utilization, and qualitatively (and quantitatively) new level of informational supporting processes. Some observed in processes have to be analysed from the perspective to be considered as a super-socio-technological system's activity. Therefore some reservations can be taken into account with respect to the existence of post-socio-technological level of dissipative processes organisation.